\documentclass[letterpaper, 10 pt, conference]{ieeeconf}  

\IEEEoverridecommandlockouts                              
\usepackage[colorlinks,linkcolor=green,citecolor=red,urlcolor=blue,bookmarks=true,hypertexnames=true]{hyperref} 
\usepackage[utf8]{inputenc}
\usepackage[T1]{fontenc}

\usepackage{graphicx} 
\usepackage{epsfig} 
\usepackage{mathptmx} 
\usepackage{mathptmx} 
\usepackage{amsmath} 
\usepackage{amssymb}  
\usepackage{booktabs}
\usepackage{xurl}
\usepackage{multirow}
\usepackage{multicol}
\usepackage{subcaption}
\usepackage{caption}  
\usepackage{array}                 
\usepackage{float}
\usepackage{balance}
\title{\LARGE \bf Traffic Signal Phase and Timing Estimation with Large-Scale Floating Car Data}

\author{Mingcheng Liao$^{1,2,\dagger}$, Zebang Feng$^{1,\dagger}$, Miao Fan$^{3,*}$, Shengtong Xu$^{4}$, Haoyi Xiong$^{5}$
\\
\\
$^{1}$NavInfo Co. Ltd.,
$^{2}$Australian National University,
$^{3}$Beijing Institute of Graphic Communication,
\\
$^{4}$Autohome Inc.,
$^{5}$Baidu Inc.
\thanks{$\dagger$Equal contribution.}
\thanks{*Corresponding author: Miao Fan (miao.fan@ieee.org), professor at Beijing Institute of Graphic Communication, senior member of IEEE.}
}

\begin{document}

\maketitle
\thispagestyle{empty}
\pagestyle{empty}

\begin{abstract}
Effective modern transportation systems depend critically on accurate Signal Phase and Timing (SPaT) estimation. However, acquiring ground-truth SPaT information faces significant hurdles due to communication challenges with transportation departments and signal installers. As a result, Floating Car Data (FCD) has become the primary source for large-scale SPaT analyses. Current FCD approaches often simplify the problem by assuming fixed schedules and basic intersection designs for specific times and locations. These methods fail to account for periodic signal changes, diverse intersection structures, and the inherent limitations of real-world data, thus lacking a comprehensive framework that is universally applicable. Addressing this limitation, we propose an industrial-grade FCD analysis suite that manages the entire process, from initial data preprocessing to final SPaT estimation. Our approach estimates signal phases, identifies time-of-day (TOD) periods, and determines the durations of red and green lights. The framework's notable stability and robustness across diverse conditions, regardless of road geometry, is a key feature. Furthermore, we provide a cleaned, de-identified FCD dataset and supporting parameters to facilitate future research. Currently operational within our navigation platform, the system analyses over 15 million FCD records daily, supporting over two million traffic signals in mainland China, with more than 75\% of estimations demonstrating less than five seconds of error.
\end{abstract}
{\keywords SPaT estimation, Large-Scale Real FCD}


\section{Introduction}
\begin{figure} 
    \centering
    \includegraphics[width=\columnwidth]{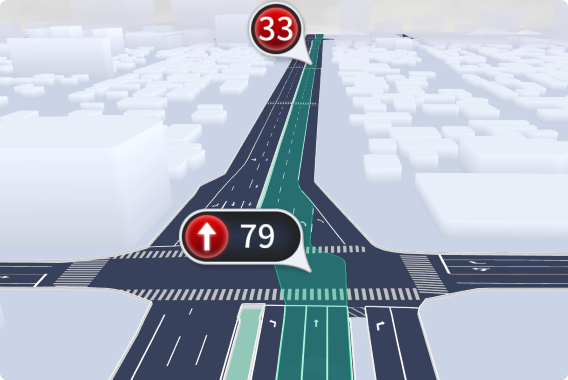}
    \caption{Our SPaT estimation framework, integrated into our navigation platform, delivers real-time traffic signal status updates to users.}
    \label{fig:framework_overview}
\end{figure}

Signal Phase and Timing (SPaT) prediction, which involves accurately estimating traffic signal phase durations (e.g., green and red cycles) and identifying corresponding Time-of-Day (TOD) schedules, represents a critical data-driven challenge within intelligent transportation systems~\cite{LEITNER2022507}. Such capability is fundamental to optimising traffic flow, mitigating congestion, enhancing safety, and enabling coordinated vehicle movements~\cite{jiangInterpretableCascadingMixtureofExperts2024, mahlerOptimalVelocityPlanningScheme2014, Asadi2011PredictiveCC}. Moreover, providing drivers with real-time SPaT data facilitates informed decision-making, leading to smoother and more predictable driving experiences, as we provide on our platform (Fig~\ref{fig:framework_overview}). However, obtaining ground-truth SPaT data directly is often infeasible due to communication barriers and the decentralised control of traffic signals among various manufacturers and technical teams. This necessitates data-driven approaches for effective SPaT prediction.

Historically, SPaT prediction research has relied on two primary data modalities. The first employs temporarily installed detectors or sensor arrays to capture detailed vehicle and intersection dynamics~\cite{genserTimetoGreenPredictionsFramework2020, genserTimetoGreenPredictionsFullyActuated2024}. The second leverages Floating Car Data (FCD), sourced either from microscopic traffic simulations (e.g., SUMO) or real-world vehicle trajectories~\cite{ugirumureraMachineLearningMethod2023, axerEstimatingSignalPhase2016}. While in-vehicle cameras can identify countdown timers, this approach lacks broad applicability due to the inconsistent presence of such displays. Although specialised sensors and simulations yield high-fidelity data, their widespread deployment, particularly across large regions like mainland China, faces significant economic and logistical hurdles. FCD can be acquired from sensors installed on vehicles, capturing both temporal information and external parameters (e.g., geographic coordinates) as well as internal parameters (e.g., vehicle speed). Through the natural movement of the vehicle fleet, this approach enables coverage of a much larger area at a lower cost. Consequently, FCD from real-world vehicles emerges as a more scalable and cost-effective solution, which can be harnessed from existing fleets for extensive traffic analysis~\cite{Garcia2014,su15010711}. Nonetheless, FCD presents its challenges: limited fleet penetration and sensor precision can result in sparse, discontinuous, and inaccurate data. Issues like suboptimal GPS positioning and random data gaps further compound the analytical complexity.

To address these inherent limitations on data sparsity, signal timing irregularities, and the prohibitive costs of infrastructure-based methods, we introduce a comprehensive framework for mining SPaT information from FCD. Our methodology focuses on robust data extraction and inference techniques to overcome the obstacles faced by conventional approaches. Specifically, the framework enhances the estimation of signal cycle lengths, TOD period boundaries, and phase durations (red and green times) within each period. We use the Fast Fourier Transform (FFT) to identify dominant cycle periodicities, combined with the Kolmogorov-Smirnov (KS) test and a tailored dispersion measure, to refine cycle length estimation and mitigate the effects of sparsity. The dispersion measure is adapted further to delineate TOD transitions precisely. Signal duration estimation employs a cross-validation strategy, improving the reliability of red and green time predictions. This end-to-end framework is currently operational on our navigation platform, providing SPaT intelligence for over two million traffic signals across major Chinese cities.

The primary contributions of this work are as follows.

\begin{itemize}
    \item \textbf{Industrial-Grade Solution:} Development of a complete pipeline, from data preprocessing to prediction, supporting a large-scale navigation platform. This system processes over 15 million FCD records daily (more than two billion km travelled) and delivers SPaT estimates for over two million signals with semi-fixed cycles in mainland China.

    \item \textbf{Novelty and Reliability:} Introduction of robust estimation techniques, including cross-validation, that ensure reliable performance even with noisy, incomplete FCD, significantly improving upon methods sensitive to data quality issues.

    \item \textbf{Reproducibility:} Public release of source code and the FCD dataset are available via OneDrive\footnote{%
  \sloppy                   
  \url{https://1drv.ms/f/s!AiPwiDNbcd1umbcgLBfuYy3IP71NEA?e=YaQPF3}%
  \par                       
}. The dataset includes over 720,000 cleaned and de-identified records from around 100 intersections, along with vision-based ground truth, which fosters transparency and facilitates further research.
\end{itemize}

\section{Related Work}

Over the past decade, research into SPaT prediction has explored various data sources, including pre-installed detectors, fixed cameras, and license plate recognition (LPR) systems. For instance, \cite{doi:10.3141/2623-06} utilised high-frequency intersection sensors to reconstruct queue dynamics and estimate signal cycle lengths. LPR data, which captures precise vehicle crossing times, has also been leveraged. Lin et al.~\cite{itr2.12198} combined density-based clustering with frequency-domain analysis, while Yao et al.~\cite{Yao31122023} developed a bi-level programming model to align observed vehicle pass-through events with theoretical cycle timings. Further integrating LPR data, Zhan et al.~\cite{ZHAN2020102660} employed Gaussian processes and dynamic linear Gaussian models to jointly estimate signal timings, queue lengths, and link travel times.

Advanced camera systems providing high-resolution trajectories represent another infrastructure-heavy approach. Zhou et al.~\cite{zhouTrafficSignalPhase2024} demonstrated a two-step method using radar-vision integrated cameras (RVIC), which employs dispersion analysis for estimating cycle length and TOD period. While effective under specific conditions (low volume and standard phasing), the need for specialised hardware limits its broad adoption. Although these infrastructure-dependent methods allow for dynamic adaptation and high accuracy, potentially overcoming assumptions of fixed timing, they incur substantial costs. Notably, analytical techniques like dispersion analysis show promise for robustness even with lower-quality data sources if adapted properly.

FCD constitutes a distinct approach to SPaT estimation, generally offering lower costs and broader coverage compared to deployed sensors. However, FCD quality is often hampered by uneven density and discontinuity arising from variable traffic and limited vehicle participation. Privacy considerations also pose challenges to data acquisition and analysis. Early FCD work includes Kerper et al.~\cite{Kerper2012}, who aggregated velocity profiles using a cloud platform for real-time estimation, and Axer and Friedrich~\cite{Axer2015, axerEstimatingSignalPhase2016}, who developed statistical methods for sparse and irregular trajectories. A standard limitation of these initial FCD studies was their primary reliance on simulation for validation.

More recent work has focused on extracting SPaT parameters from real-world FCD sources. Chuang~\cite{Chuang2015} used stop-to-go transitions in coarse-grained FCD as indicators of phase timing. Du et al.~\cite{duSignalTimingParameters2019a} exploited high-resolution FCD from ride-hailing vehicles to estimate a comprehensive set of parameters (cycle length, TOD points, phase sequence, green times) by analysing cyclic patterns. Yu et al.~\cite{yuLearningTrafficSignal2016} applied an approximate greatest common divisor (AGCD) model to low-sampling-rate taxi GPS data. However, such data inherently risks missing critical events or yielding inconsistent features, which can potentially impact accuracy. Utilising transit bus GPS data, Fayazi and Vahidi~\cite{fayaziTrafficSignalPhase2015,7323843} proposed methods with enhanced filtering for sparsity and noise. However, potential biases from unique bus behaviours and incomplete handling of queue delays may limit applicability, especially at congested locations.

In summary, while previous studies have utilized diverse data sources like specialized sensors~\cite{doi:10.3141/2623-06}, LPR systems~\cite{itr2.12198, Yao31122023, ZHAN2020102660}, advanced cameras~\cite{zhouTrafficSignalPhase2024}, or various forms of FCD~\cite{Kerper2012, Axer2015, axerEstimatingSignalPhase2016, Chuang2015, duSignalTimingParameters2019a, yuLearningTrafficSignal2016, fayaziTrafficSignalPhase2015, 7323843}, they often face limitations regarding cost, scalability, reliance on specific infrastructure, assumptions about signal behavior, or sensitivity to the quality and sparsity of real-world FCD. This work contrasts with prior efforts by proposing a comprehensive, end-to-end framework specifically designed for large-scale, potentially imperfect FCD. Our approach uniquely combines techniques like FFT with KS validation and adaptive dispersion analysis to ensure robustness, without assuming fixed schedules or simplified intersection geometries, and addresses the practical challenges of real-world deployment.

\section{Method}
The methodology begins with the data processing module, where raw FCD is cleaned to remove noise and mapped to the corresponding intersections. Multi-day data is then superimposed to enhance signal consistency and robustness. The cycle length estimation model analyses the preprocessed FCD to predict the signal cycle length. Based on this prediction and the FCD data, the TOD estimation module and the signal duration estimation module further analyse the data to produce their respective outputs.

\subsection{Data Preprocessing}
Raw FCD is recorded based on timestamps and includes the vehicle's current properties, collected by sensors installed on the vehicles at a sampling rate of every 3 seconds. However, the data may contain location biases or missing points due to poor GPS signals caused by environmental or positional factors, like tunnels, urban canyons, or signal obstructions. Therefore, initial preprocessing is required to filter noise and extract meaningful information from the raw FCD. Our pipeline begins with data cleaning, filtering out abnormal trajectories characterised by obvious errors, noise, or unrealistic driving patterns. We then segment the valid trajectories and perform map matching against our digital road network. Simultaneously, we incorporate corresponding road information to calculate essential metrics, such as the vehicle's distance to the stop line and intersection passing time.

\begin{figure}[ht]
    \centering
    \includegraphics[width=\columnwidth]{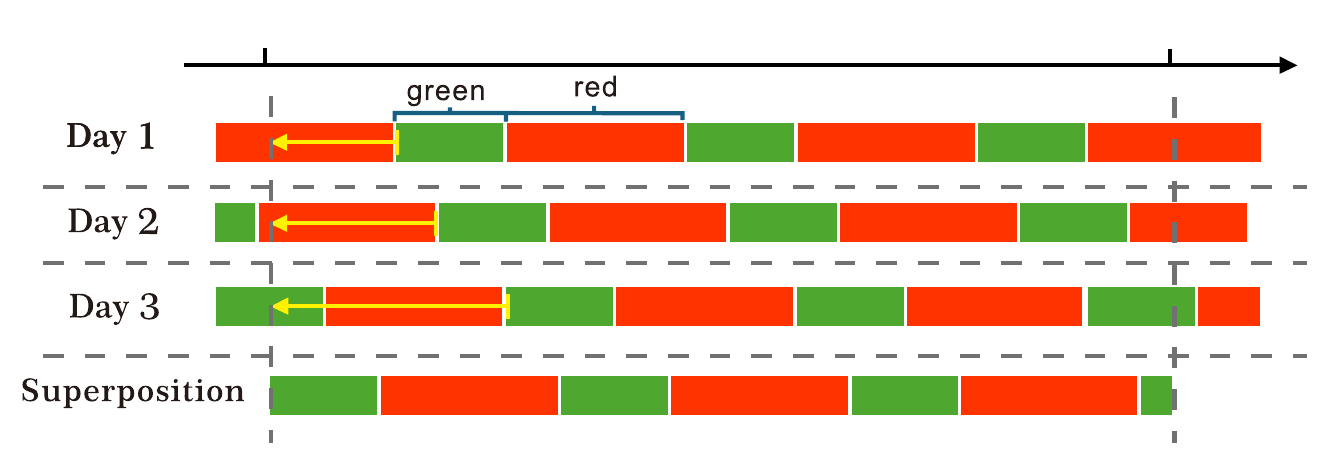}
    \caption{FCD for the same intersection, illustrating overlapping trajectories across multiple days concerning the first stop line crossing event.}
    \label{fig: superposition}
\end{figure}
To address the inherent temporal sparsity of FCD, we employ a multi-day data superposition strategy to process sufficient data for robust analysis, as illustrated in Fig~\ref{fig: superposition}. This strategy assumes, based on standard traffic management practices, that signal cycle lengths remain consistent across weekdays and, separately, across weekend days. The primary feature extracted for SPaT estimation is the start time of the vehicle, specifically, the timestamp of the stop-to-start transition near the intersection. Analysing raw start times typically yields poor results due to high sensitivity to variations in stopping position relative to the stop line. Therefore, a calibration process is crucial. This calibration uses a proprietary link ranking system and associated parameters derived from extensive historical data analysis, based on linear regression factors such as time of day, road type, city, importance, width, and category. Conceptually, this calibration adjusts the recorded start times of vehicles stopping further back in the queue to align with those stopping closer to the front.

Given the sparse nature of FCD, valid data points may occur sporadically throughout the day. Attempting to infer the signal phase solely from the absolute timestamp can lead to significant errors due to the complexity of real-world signal scheduling. To mitigate this, our method aggregates data from all relevant days (e.g., weekdays) within a month into 1-hour analysis windows. An alignment process is performed within each window using the first second of the window as a temporal reference. For each day contributing data to that window, the start time of the first valid trajectory is identified. The time difference between this first start time and the reference time of the window is calculated. This offset is then applied to normalise the timestamps of all trajectories from that specific day within that 1-hour window, effectively synchronising the relative start times across different days.

\subsection{Estimating Cycle Length}
\begin{figure}[htbp]
  \centering
  \begin{subfigure}[b]{\columnwidth}
    \centering
    \includegraphics[width=\columnwidth]{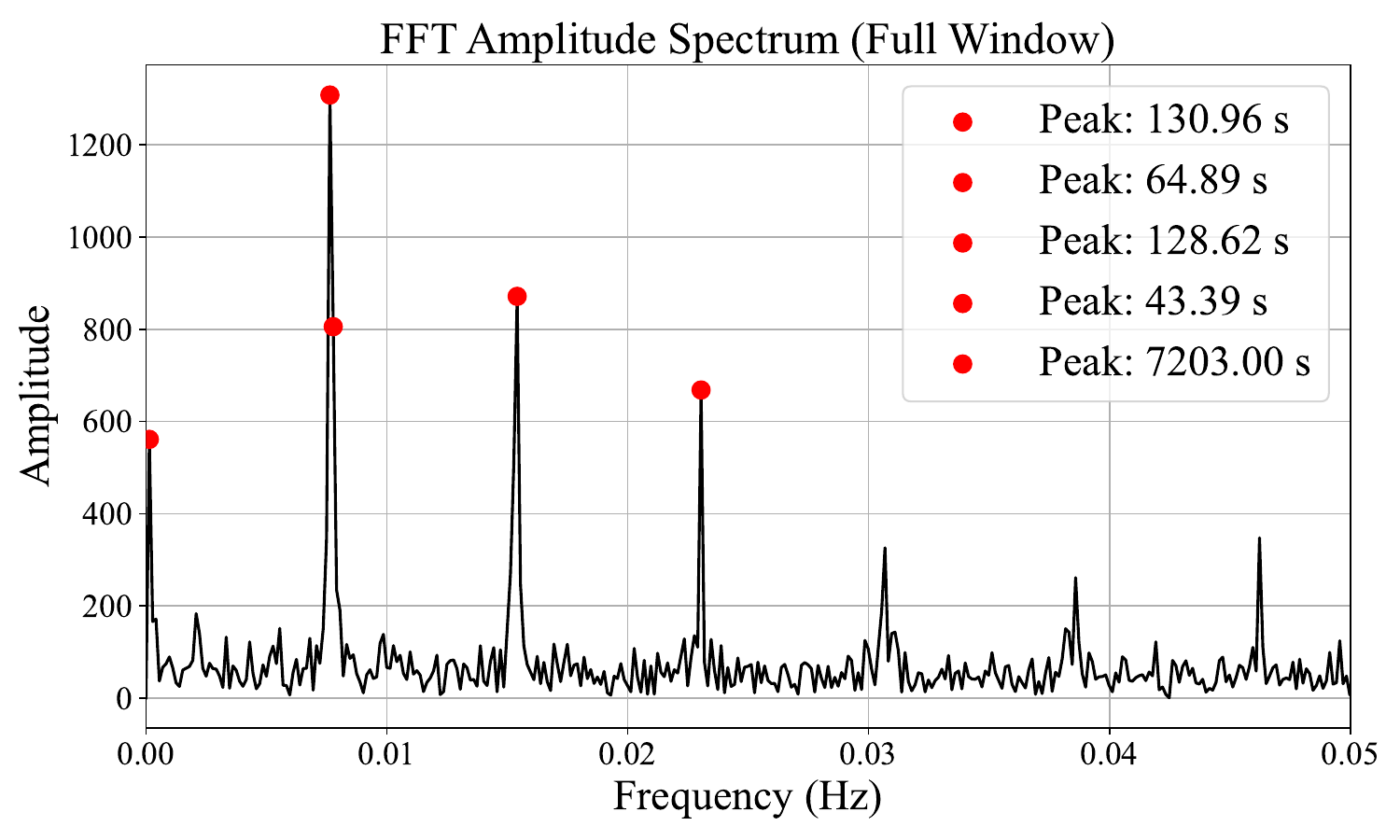}
    \caption{Candidate frequencies identified via FFT.}
    \label{fig:fft}
  \end{subfigure}
  
  \vspace{1em} 
  
  \begin{subfigure}[b]{\columnwidth}
    \centering
    \includegraphics[width=\columnwidth]{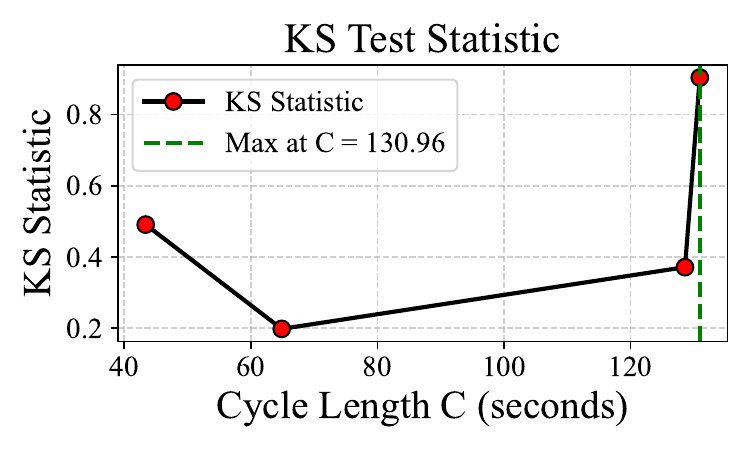}
    \caption{Verification via the KS test.}
    \label{fig:ks}
  \end{subfigure}
  
  \caption{Candidate frequencies are introduced via FFT~\ref{fig:fft} and then verified using the KS test ~\ref{fig:ks}.}
  \label{fig:fft_ks}
\end{figure}

With information on the relative starting times of all vehicles, the frequency data can be mapped onto a timeline and used to analyse the cycle length. The Fast Fourier Transform (FFT) converts a signal from the time domain to the frequency domain. It is a traditional and effective method for extracting the cycle length. We propose using the FFT to identify initial candidate periods in the data, as illustrated in Fig.~\ref{fig:fft}. The FFT algorithm is very robust when the FCD data is continuous and dense~\cite{itr2.12198}; however, due to the sparsity of the data, FFT struggles to distinguish the actual cycle length from frequency peaks associated with its half-wavelength and multiple wavelengths. 

To address this issue, we combine FFT with additional optimisation techniques. We map all the data onto the candidate cycles proposed by the FFT,

\begin{equation}
    t_{\text{map}} = t - \mathrm{ROUND}\left(\frac{t}{C}\right) \cdot C,
\end{equation}
where $C$ is the candidate cycle length, $\text{t}$ represents the actual Unix timestamp, and $t_{\text{map}}$ is the timestamp mapped to the candidate cycle length.

If a candidate cycle represents the primary component of the data, its distribution is expected to be approximately normal rather than uniform. Therefore, we propose using the Kolmogorov–Smirnov (KS) test to assess the degree of normality, as it robustly distinguishes the true cycle from its halves or multiples. The one-sample KS test compares the difference between the empirical distribution function and the cumulative distribution function (CDF) of a theoretical distribution, as illustrated in the following equation:

\begin{equation}
    D = \mathrm{MAX}(F_n(t_{\text{map}}) - G_m(t_{\text{map}})),
\end{equation}
where $F_n(t_{\text{map}})$ represents the starting frequency at the mapped time, $G_m(t_{\text{map}})$ is the CDF of the theoretical distribution, and $D$ is the KS statistic, which quantifies the extent of the differences between the two distributions. In the case shown in Fig.~\ref{fig:fft_ks}, we compare the distribution of the mapped frequency data with a uniform distribution and select the candidate cycle corresponding to the highest Kolmogorov–Smirnov (KS) statistic.

\subsection{Estimating of TOD}
The TOD searching algorithm typically achieves minute-level accuracy on our dataset, as actual schedule change timings generally coincide with fluctuations in traffic volume and are supported by sufficient data during these periods. Moreover, the algorithm presumes that the traffic schedule remains unchanged late at night when vehicle presence is minimal.

\begin{figure}[ht]
    \centering
    \includegraphics[width=\columnwidth]{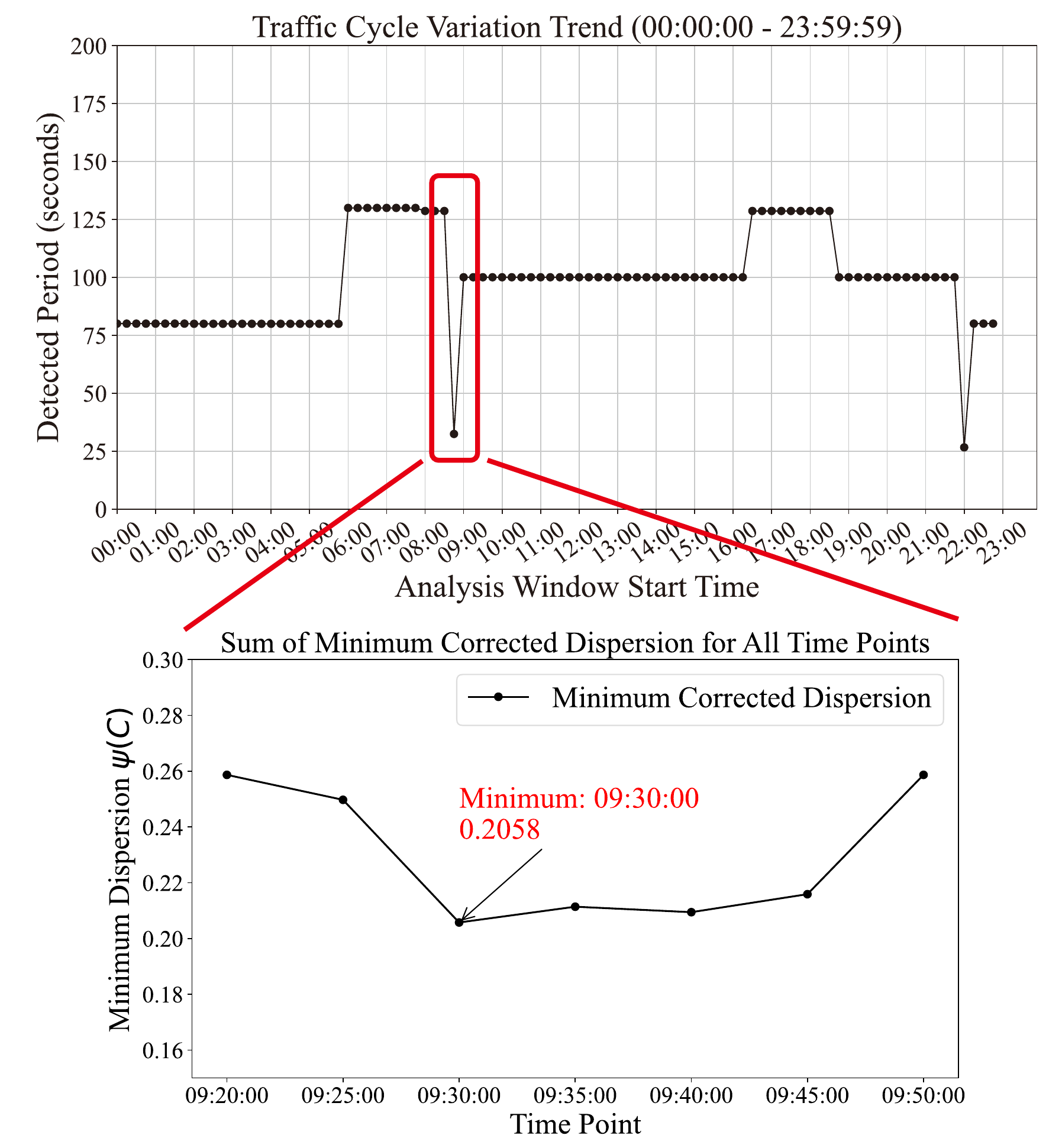}
    \caption{Sliding window search for potential period-switch range.}
    \label{FFT-convolution}
\end{figure}

The algorithm addresses the problem in two steps. First, a quick search is performed using FFT. Then, the concept of dispersion is introduced for precise positioning. While FFT operates stably when the traffic volume is stable and continuous, it tends to yield fluctuating and unreasonable period estimates when a window spans two schedule changes, as illustrated in Fig~\ref {FFT-convolution}. The FFT window is configured to 1 hour with a step size of 15 minutes, ensuring that the algorithm receives sufficient data support to detect timing changes. When a clear schedule change or fluctuation occurs between two windows, the algorithm can determine the approximate time range during which the change takes place.

\begin{figure}[ht]
    \centering
    \includegraphics[width=\columnwidth]{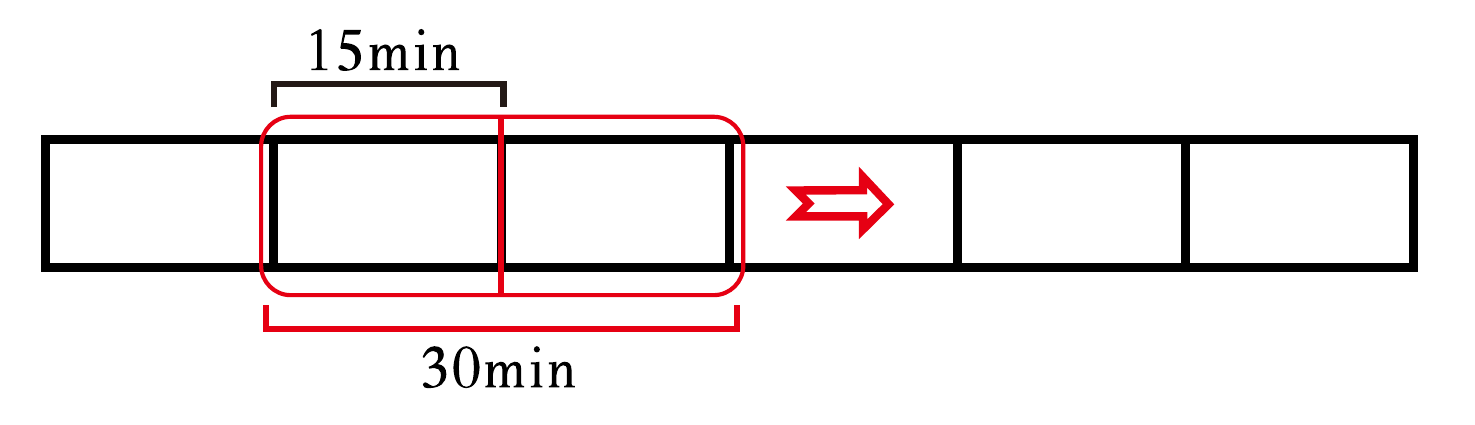}
    \caption{A two-section sliding window approach performing refined search during potential period-switch intervals.}
    \label{fig: search}
\end{figure}

Zhou et al.~\cite{zhouTrafficSignalPhase2024} introduced a dispersion method for precisely verifying the potential cycle length. We have redesigned this approach by adapting the dispersion definition to our dataset and incorporating an initial screening step using the FFT. This modification prevents the dispersion from escalating excessively when handling large datasets, thereby maintaining control over the subsequent penalised term, where the original dispersion is defined as, 
\begin{equation}
    \psi_p(C)=\frac{\sqrt{\frac{\sum_{j=1}^{n_{g s}} \delta_{i^*, j}{ }^2}{n_{gs}}}}{C},
\end{equation}
where the $n_{g s}$ is the sequence containing all the mapped green start times, and $\delta_{i^*, j}$ represents the time differences between the other green start times and the most common green start time. Since dispersion is sensitive to the half-wavelength, we also need to add a penalty to the dispersion to prevent it from converging to an incorrect optimal value. For our data, the penalty is defined as:
\begin{equation}
    \text{P} = w(1-C/C_{\text{max}})^2,
\end{equation}
where $w$ denotes the penalty weight, and $C_{\text{max}}$ denotes the theoretical upper bound of the cycle length. For our data, we set $w = 0.1$ and $C_{\text{max}} = 600$ seconds, implying that the expected maximum red phase duration is no longer than 10 minutes.

The precise search is conducted using two sliding windows, as illustrated in Fig~\ref{fig: search}. The midpoints of these windows move within the range defined by the FFT slip search, aiming to identify the optimal split point that divides two signals with distinct frequencies. As the windows slide, we select the step at which the two windows yield different estimated cycle lengths and record the corresponding maximum degree of dispersion. When the split is successful, both windows exhibit low dispersion; thus, the step with the lowest maximum dispersion is identified as the actual splitting point.

\subsection{Estimating The Duration of The Signal Durations}
Compared to previous estimations, signal duration prediction relies heavily on both the baseline phase estimate and the quality of the FCD. Consequently, poor quality data inevitably results in lower accuracy. The challenge is to derive reasonable estimations from such data using a simple method, even when ground truth values are unavailable, while ensuring reliable and non-spurious outcomes.

\begin{figure}[ht]
    \centering
    \includegraphics[width=\columnwidth]{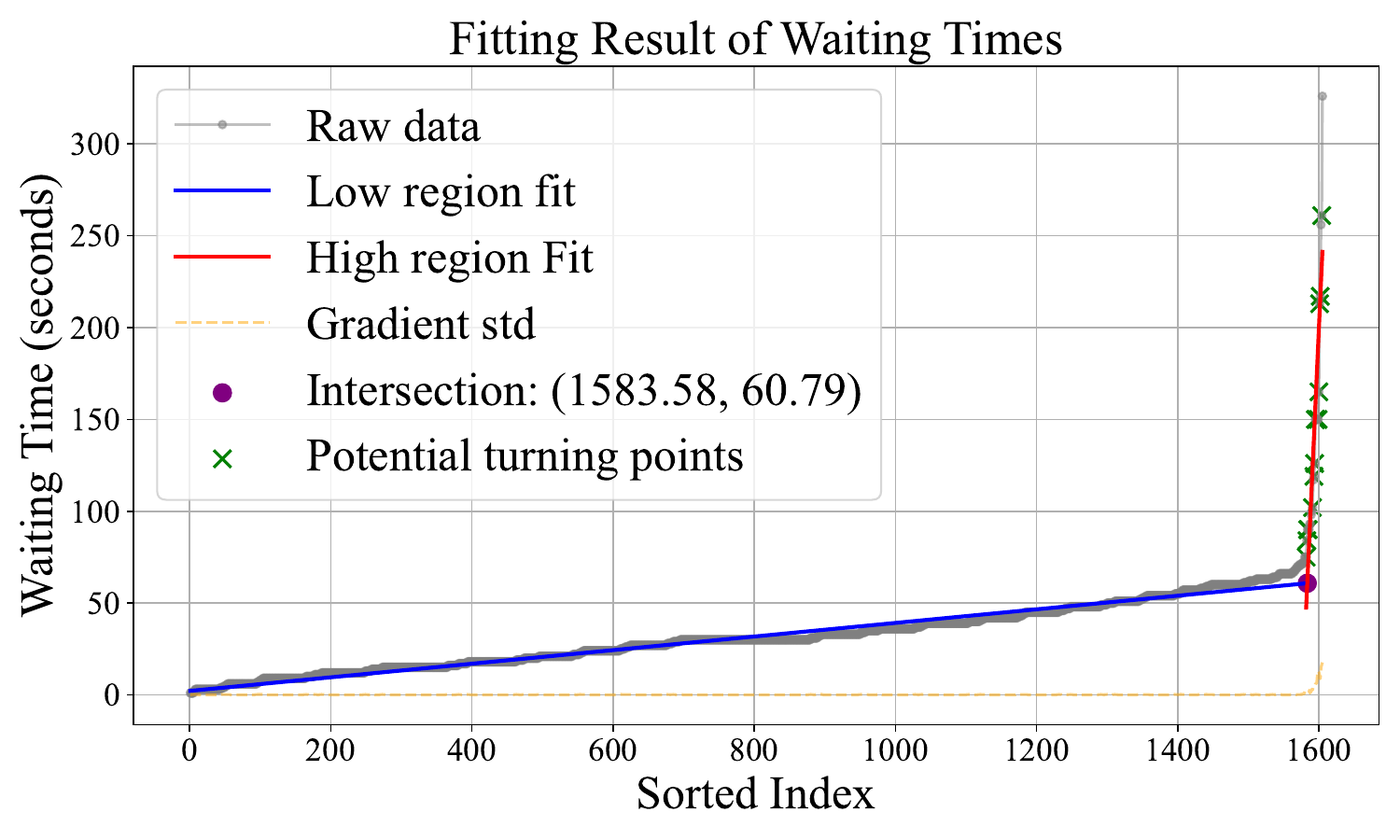}
    \caption{By plotting the sorting index against its stopping duration, the inflexion point is determined using the standard deviation of gradient changes computed over a sliding window.}
    \label{fig: red_duration}
\end{figure}

Our signal estimation approach is based on the duration that vehicles remain stopped. Analysis of the FCD reveals a range of waiting behaviours. While the majority of vehicles exhibit single-round waiting times shorter than the actual red duration, some experience full-stop periods or rear-start delays that cause their waiting times to approximate the actual red duration closely. Moreover, a subset of vehicles display significantly longer waiting times, possibly due to activities such as nearby parking or delayed start. Consequently, ranking the waiting times within a phase typically yields a distribution, as illustrated in Fig~\ref {fig: red_duration}.

Based on this analysis, the actual red period is optimally defined at the inflexion point of the waiting time distribution. To identify this inflexion, we propose a gradient analysis method. Specifically, We first sort all observed waiting times $t_i$ into ascending order $t_{(1)}\le t_{(2)}\le\cdots\le t_{(N)}$ and assign them integer indices $i=1,\dots,N$. A discrete gradient series is then computed by 
\begin{equation}
      g_i = t_{(i+1)} - t_{(i)},\quad i=1,\dots,N-1,
\end{equation}
using a one-sided finite difference at the boundaries. To characterise local fluctuation in the gradient at each position $i$, we compute a sliding-window sample standard deviation 
\begin{equation}
      \sigma_i = 
    \mathrm{std}\!\bigl(g_{\,\max(1,i-w)} , \dots, g_i, \dots, g_{i+w}\bigr),
\end{equation}
with window size $w$. A global dynamic threshold is then set as 
\begin{equation}
      \theta = \alpha \cdot \frac1{N}\sum_{j=1}^N \sigma_j,
\end{equation}
where $\alpha$ is a tunable `pressure' factor. We scan $i$ and declare a turning point wherever $|g_i - g_{i-1}| > \theta$. The first such index, $i^*$, is taken as our primary inflection point. To avoid local noise around $i^*$, we exclude all points in $[i^*-W,i^*+W]$ with $W=20$, then split the remaining sorted data into 'low-waiting' segment $i<i^*$ and `high-waiting' segment $i>i^*$. Finally, we perform separate linear regressions
\begin{equation}
    L_{\rm low}(i)=a_\ell\,i+b_\ell,\quad
L_{\rm high}(i)=a_h\,i+b_h
\end{equation}
on the two segments, and compute their analytical intersection
\begin{equation}
i^*_{\rm pred} = \frac{b_h - b_\ell}{a_\ell - a_h},\quad
t^*_{\rm pred} = a_\ell\,i^*_{\rm pred} + b_\ell,
\end{equation}
which yields our predicted red-phase boundary index and its corresponding waiting time.

\begin{figure}[!h]
    \centering
    \includegraphics[width=\columnwidth]{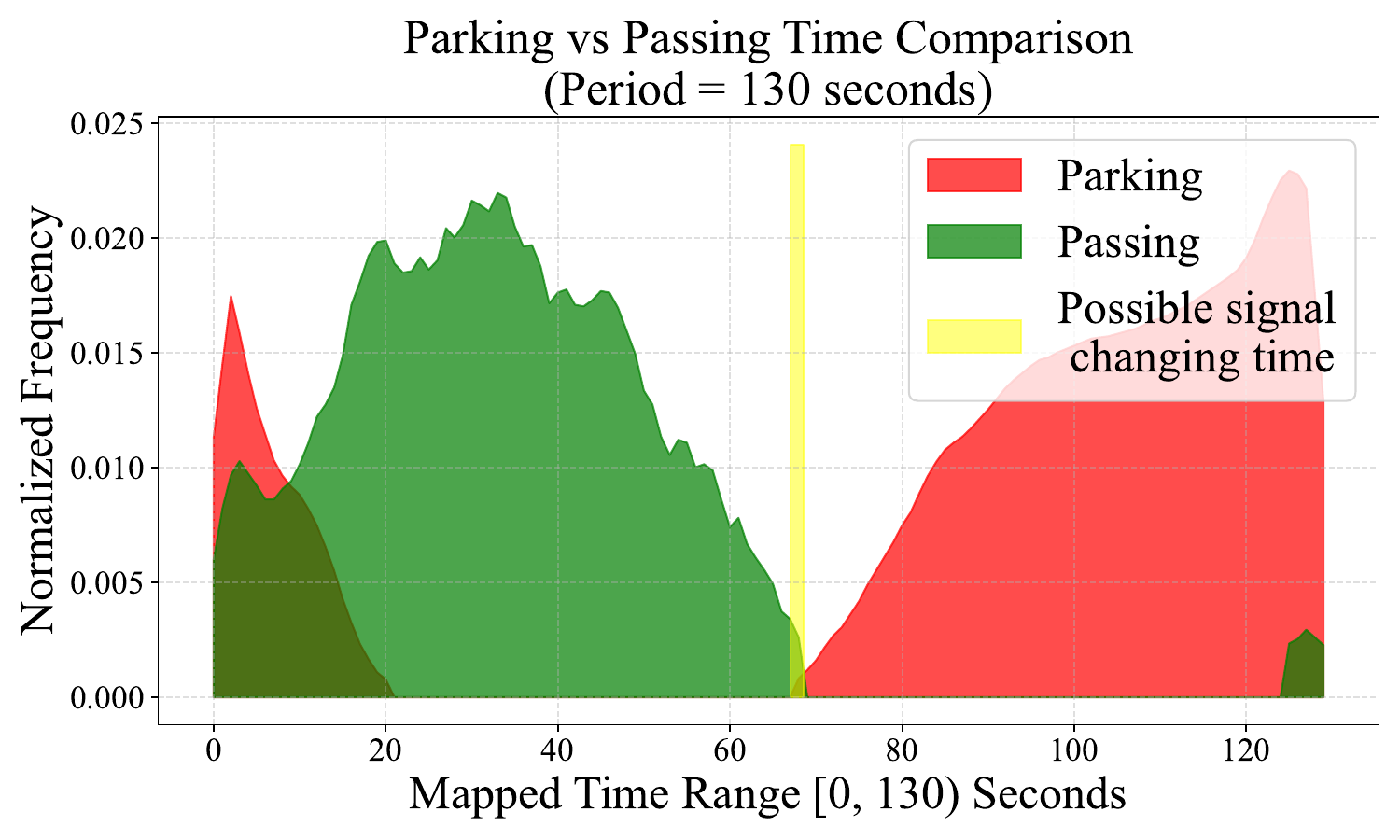}
    \caption{By overlaying the vehicle passing frequency and stopping frequency, the red-green phase is defined at 0 seconds, and the yellow area indicates the potential green-red switching time.}
    \label{fig: red-green}
\end{figure}

However, our method may yield predictions with significant deviations in rare cases. We propose a vote confirmation strategy to detect anomalies without ground-truth signal durations. This approach maps the frequency of vehicles passing the stop line and the count of waiting cars at each second onto the estimated period, followed by normalisation. As shown in Fig~\ref{fig: red-green}, superimposing data from different days based on the start time of the vehicles confirms that the 0-second mark corresponds to the transition from red to green. Although varying traffic conditions can complicate the correlation between vehicle stop timing and the onset of the red signal, we can verify that the green-to-red transition occurs within the interval between the lowest count of waiting vehicles and the point where the red and green signals crossover. Therefore, if the prediction falls outside this interval, it is recalled and classified as faulty. Furthermore, in the actual situation, where applicable, we plan to perform additional confirmation based on standard traffic rules using the method described in~\cite{duSignalTimingParameters2019a}.

\section{Experiment}
\subsection{Dataset}
Our empirical evaluation relies on three primary data components: the processed FCD data, a visually verified ground truth dataset, and a set of calibration parameters.

The processed FCD contains detailed trajectory information, including the city of origin, road importance level, intersection identifier (ID), entering direction, steering data, entry and exit Unix timestamps, intersection travel time, and specific stopping behavior details (waiting time, distance to the stop line, and vehicle stop timestamp). The ground truth dataset, derived from visual inspection or known signal plans, provides corresponding intersection IDs, timestamps, and true SPaT parameters, such as cycle length and the durations of individual phases (red or green). Finally, the calibration parameter set contains coefficients used to model and compensate for queue-induced start delays, parameterised by city, road level, time interval, and specific linear calibration values derived from historical data.

This study used a large-scale, realistic FCD collection spanning over 10,000 intersections in major Chinese cities, including Beijing, Shanghai, and Shenzhen. This dataset encompasses a diverse range of road structures and traffic flow conditions, ensuring a comprehensive evaluation. Table~\ref{table: datasets} summarises the data distribution across these cities. Beijing contributes data from 5,241 intersections, while Shanghai and Shenzhen provide data from 2,987 and 2,076 intersections, respectively. On average, each analysed intersection section includes approximately 710 valid trajectories.




\begin{table}
    \centering
    \caption{Statistics information of the datasets.}
    \label{table: datasets}
    \begin{tabular}{lrrr}
    \toprule
     \textbf{Item} & \textbf{Beijing} & \textbf{Shanghai} & \textbf{Shenzhen} \\
    \midrule
    Intersections &  5,241 & 2,987 & 2,076 \\
    Trajectories & 3.9 M & 2.1 M & 1.1 M \\
    \bottomrule
    \end{tabular}
\end{table}

\subsection{Experiment Setting}

\begin{table*}[!t]
\centering
\caption{Estimation of signal cycle length in different time and locations.}
\label{tab:cycle_length}
\begin{tabular*}{\textwidth}{@{\extracolsep{\fill}}lrrrrrr@{}}
    \toprule
    \multicolumn{1}{c}{\textbf{Dataset}} & \multicolumn{6}{c}{\textbf{Method}} \\
    \midrule
    & \multicolumn{2}{c}{FFT Only} & \multicolumn{2}{c}{MFAGCD} & \multicolumn{2}{c}{Our method} \\
    \cmidrule(lr){2-3}\cmidrule(lr){4-5}\cmidrule(lr){6-7}
    & ACC-5 & ACC-3 & ACC-5 & ACC-3 & ACC-5 & ACC-3 \\
    \midrule
    Overall  & 0.805 & 0.746 & 0.863 & 0.818 & 0.913 & 0.815 \\
    \midrule
    rush hours (Beijing) &  0.875 &  0.823 & 0.884 &  0.782 &  0.941& 0.827 \\
    off-peak hours (Beijing) &  0.739 &  0.697 &  0.810 & 0.724 & 0.836 & 0.767 \\
    \midrule
    rush hours (Shanghai) & 0.816 & 0.737 & 0.879 & 0.814 & 0.899 & 0.829 \\
    off-peak hours (Shanghai) & 0.768 & 0.710 & 0.789 & 0.722 & 0.829 & 0.737 \\
    \midrule
    rush hours (Shenzhen) & 0.746 & 0.688 & 0.792 & 0.721 & 0.787 & 0.723 \\
    off-peak hours (Shenzhen) & 0.692 & 0.621 & 0.751 & 0.703 & 0.711 & 0.697 \\
    \bottomrule
\end{tabular*}
\end{table*}

\begin{table*}[!t]
\centering
\caption{Estimation of signal duration in different time and locations.}
\label{tab:signal_phase}
\begin{tabular*}{\textwidth}{@{\extracolsep{\fill}}lrrrrrr@{}}
    \toprule
    \multicolumn{1}{c}{\textbf{Dataset}} & \multicolumn{6}{c}{\textbf{Method}} \\
    \midrule
    & \multicolumn{2}{c}{ADJD} & \multicolumn{2}{c}{Queue release} & \multicolumn{2}{c}{Our method} \\
    \cmidrule(lr){2-3}\cmidrule(lr){4-5}\cmidrule(lr){6-7}
    & ACC-5 & ACC-3 & ACC-5 & ACC-3 & ACC-5 & ACC-3 \\
    
    \midrule
    Overall  & 0.825 & 0.755 & 0.779 & 0.721 & 0.848 & 0.769 \\
    \midrule
    rush hours (Beijing) & 0.840 & 0.782 & 0.801 & 0.745 & 0.853 & 0.776 \\
    off-peak hours (Beijing) & 0.783 & 0.697 & 0.746 & 0.665 & 0.782 & 0.701 \\
    \midrule
    rush hours (Shanghai) & 0.863 & 0.753 & 0.786 & 0.724 & 0.881 & 0.792 \\
    off-peak hours (Shanghai) & 0.739 & 0.691 & 0.763 & 0.710 & 0.848 & 0.724 \\
    \midrule
    rush hours (Shenzhen) & 0.763 & 0.658 & 0.739 & 0.695 & 0.787 & 0.715 \\
    off-peak hours (Shenzhen) & 0.678 & 0.654 & 0.688 & 0.614 & 0.691 & 0.635 \\
    \bottomrule
\end{tabular*}
\end{table*}

\subsubsection{Metrics}
We evaluated the accuracy of cycle length and signal duration estimations using data segmented into 30-minute intervals extracted from multiple time slices between 6:00 AM and 9:00 PM. Two standard metrics, Acc-3 and Acc-5, were employed. These metrics denote the proportion of predictions whose absolute error, compared to the ground truth, is within 3 seconds and 5 seconds, respectively. Formally, given a set of predicted values $\hat{y}_i$ and corresponding ground truth values $y_i$, Acc-k is defined as:
\begin{equation}
    \text{ACC-k} = \frac{1}{N} \sum_{i=1}^{N} \mathbf{1}(|\hat{y}_i - y_i| \leq k),
\end{equation}
where $k \in \{3, 5\}$, $N$ is the number of samples, and $\mathbf{1}(\cdot)$ is the indicator function. The accuracy of TOD period identification was evaluated implicitly through its impact on cycle length estimation within those periods. Additionally, we used a recall rate metric (applied to the complete test set) to quantify the framework's capability to produce reasonable predictions and identify instances potentially compromised by severe data limitations. 


\subsubsection{Implement detail}
For cycle length estimation, our implementation generates five candidate periods using FFT, with the KS test providing confidence levels for selection. TOD period detection involves an initial coarse search using FFT with a 1-hour sliding window and a 15-minute step size, followed by a fine-grained search using the dispersion measure with a 30-minute window and a 5-minute step size. Signal duration estimation relies on gradient analysis over 3-second intervals ($w = 3$); an adjacent gradient change exceeding 10 times ($\alpha = 10$) the local standard deviation (within its calculation window) signifies a potential signal phase transition point.

\subsection{Method of Comparison}
We compared the performance of our proposed framework against several established or representative baseline methods using our large-scale dataset. The online experiment results are presented subsequently.

\subsubsection{FFT only} For the aggregated vehicle starting data, this approach applies FFT directly to analyse cycle lengths without incorporating the KS test for additional verification.

\subsubsection{Most-frequent approximate greatest common divisor (MFAGCD)~\cite{yuLearningTrafficSignal2016}} This method formulates cycle length estimation as an AGCD problem applied to a set of sparse green-start times extracted from taxi GPS data. Essentially, it identifies the most frequently occurring "approximate divisor" among the inter-green time intervals.

\subsubsection{All-direction joint determination (ADJD)} On the same dataset, this method simultaneously estimates green durations for all phases by integrating green-start and crossover times across multiple directions, thereby leveraging their correlations to improve both accuracy and robustness.

\subsubsection{Queue release analysing method~\cite{duSignalTimingParameters2019a}} This approach utilises floating car data to identify the instances when vehicles cross the stop line during a green phase. By clustering these crossing times, both the effective green duration and cycle length are calculated. Although this method bears similarity to our verification strategy in signal timing estimation, it directly analyses queue release behaviour to derive the predictions.

\subsection{Offline Experiment}
Tables~\ref{tab:cycle_length} and~\ref{tab:signal_phase} present the estimation results for signal cycle length and duration across different time segments and locations in our large-scale dataset. In summary, our method outperforms baseline approaches in cycle length prediction and exhibits superior robustness in signal duration estimation, particularly when handling imperfect data and varying environmental conditions.

In comparison, relying solely on FFT results tends to be less robust when data quality is compromised, often leading to predictions that consider the entire time section as the cycle length. Although the MFAGCD method achieves similar performance, it depends on the assumption that the lower and upper bounds of the traffic signal cycle are known, a premise that may not hold in real-world scenarios and can consequently result in inaccurate predictions. In contrast, our approach eliminates the need for such assumptions by analysing candidate periods generated from the FFT, thereby increasing the likelihood of accurate estimation.

For signal duration estimation, both the ADJD and queue release analysis methods demonstrate lower performance on our dataset. A plausible explanation is that these methods rely exclusively on vehicle passage times at intersections to capture signal phase changes, thereby neglecting factors such as vehicle localisation errors, inaccuracies in stop line localisation, and variations in driver behaviour. Nevertheless, since the overall trend in vehicle passage times remains generally reliable, we employed these methods for confirmation rather than as the primary source for direct estimation.

\subsection{Online Experiment}

\begin{table}[H]
    \centering
    \caption{The Online Performance.}
    \label{table: online}
    \begin{tabular}{lrrr}
    \toprule
     \textbf{Estimation} & \textbf{ACC-5} & \textbf{ACC-3} \\
    \midrule
    \textit{Cycle length} & 0.827 & 0.734  \\
    \textit{Signal phase} & 0.764 & 0.696  \\
    \bottomrule
    \end{tabular}
    \label{table: online}
\end{table}
We implemented an automated data pipeline for online deployment using Spark\footnote{https://spark.apache.org/}, an open-source distributed computing framework. 
We conducted road tests at 10,000 randomly sampled intersections across mainland China to evaluate the online performance. Table~\ref{table: online} presents the results using Acc-5 and Acc-3 metrics for both cycle length and signal phase estimations. The system achieved $82.7\%$ accuracy within $\pm 5$ seconds and $73.4\%$ within $\pm 3$ seconds for cycle length prediction, while signal phase estimation achieved $76.4\%$ (Acc-5) and $69.6\%$ (Acc-3). These results indicate that the proposed method maintains robust and reliable performance even in large-scale, real-world deployment scenarios with heterogeneous data quality and road conditions

\section{Conclusion} 
This paper proposes a comprehensive and innovative framework for SPaT information prediction using real, large-scale FCD without relying on prior assumptions such as fixed signal ranges or specific road structures. Our approach covers all stages, from data preprocessing to information estimation, and is evaluated against state-of-the-art methods using a large-scale real-world dataset. To facilitate reproducibility and further research, we have also released de-identified data. The evaluation results demonstrate that our framework exhibits superior adaptability to imperfect FCD while maintaining high accuracy in estimating traffic signal cycle lengths, TOD periods, and signal durations, achieving an overall accuracy of approximately 86\%. Additionally, online evaluations indicate that the method performs robustly under a wide range of conditions. 

Looking forward, our future research will prioritize improving data collection strategies to enhance data quality and prediction accuracy. Additionally, we will explore advanced deep-learning and reinforcement-learning approaches to strengthen the framework's adaptability, particularly under sparse or complex traffic conditions. Integrating Vehicle-to-Everything (V2X) communication is another promising direction, as it can provide richer, real-time traffic information, significantly enhancing SPaT prediction performance. Finally, we aim to refine model parameters and methods to better accommodate diverse road structures and traffic control strategies.
\balance
\bibliographystyle{IEEEtran}
\bibliography{IEEEexample}

\begin{thebibliography}{10}
\providecommand{\url}[1]{#1}
\csname url@rmstyle\endcsname
\providecommand{\newblock}{\relax}
\providecommand{\bibinfo}[2]{#2}
\providecommand\BIBentrySTDinterwordspacing{\spaceskip=0pt\relax}
\providecommand\BIBentryALTinterwordstretchfactor{4}
\providecommand\BIBentryALTinterwordspacing{\spaceskip=\fontdimen2\font plus
\BIBentryALTinterwordstretchfactor\fontdimen3\font minus \fontdimen4\font\relax}
\providecommand\BIBforeignlanguage[2]{{%
\expandafter\ifx\csname l@#1\endcsname\relax
\typeout{** WARNING: IEEEtran.bst: No hyphenation pattern has been}%
\typeout{** loaded for the language `#1'. Using the pattern for}%
\typeout{** the default language instead.}%
\else
\language=\csname l@#1\endcsname
\fi
#2}}

\bibitem{LEITNER2022507}
D.~Leitner, P.~Meleby, and L.~Miao, ``Recent advances in traffic signal performance evaluation,'' \emph{Journal of Traffic and Transportation Engineering (English Edition)}, vol.~9, no.~4, pp. 507--531, 2022.

\bibitem{jiangInterpretableCascadingMixtureofExperts2024}
W.~Jiang, J.~Han, H.~Liu, T.~Tao, N.~Tan, and H.~Xiong. Interpretable {{Cascading Mixture-of-Experts}} for {{Urban Traffic Congestion Prediction}}.

\bibitem{mahlerOptimalVelocityPlanningScheme2014}
G.~Mahler and A.~Vahidi, ``An {{Optimal Velocity-Planning Scheme}} for {{Vehicle Energy Efficiency Through Probabilistic Prediction}} of {{Traffic-Signal Timing}},'' vol.~15, no.~6, pp. 2516--2523.

\bibitem{Asadi2011PredictiveCC}
B.~Asadi and A.~Vahidi, ``Predictive cruise control: Utilizing upcoming traffic signal information for improving fuel economy and reducing trip time,'' \emph{IEEE Transactions on Control Systems Technology}, vol.~19, pp. 707--714, 2011.

\bibitem{genserTimetoGreenPredictionsFramework2020}
A.~Genser, L.~Ambuhl, K.~Yang, M.~Menendez, and A.~Kouvelas, ``Time-to-{{Green}} predictions: {{A}} framework to enhance {{SPaT}} messages using machine learning,'' in \emph{2020 {{IEEE}} 23rd {{International Conference}} on {{Intelligent Transportation Systems}} ({{ITSC}})}.\hskip 1em plus 0.5em minus 0.4em\relax IEEE, pp. 1--6.

\bibitem{genserTimetoGreenPredictionsFullyActuated2024}
A.~Genser, M.~A. Makridis, K.~Yang, L.~Ambühl, M.~Menendez, and A.~Kouvelas, ``Time-to-{{Green Predictions}} for {{Fully-Actuated Signal Control Systems With Supervised Learning}},'' vol.~25, no.~7, pp. 7417--7430.

\bibitem{ugirumureraMachineLearningMethod2023}
J.~Ugirumurera, J.~Severino, E.~A. Bensen, Q.~Wang, and J.~Macfarlane. A {{Machine Learning Method}} for {{Predicting Traffic Signal Timing}} from {{Probe Vehicle Data}}.

\bibitem{axerEstimatingSignalPhase2016}
S.~Axer and B.~Friedrich, ``Estimating signal phase and timing for traffic actuated intersections based on low frequency {{Floating Car Data}},'' in \emph{2016 {{IEEE}} 19th {{International Conference}} on {{Intelligent Transportation Systems}} ({{ITSC}})}.\hskip 1em plus 0.5em minus 0.4em\relax IEEE, pp. 2059--2064.

\bibitem{Garcia2014}
A.~Garcia-Castro and A.~Monzón, ``Using floating car data to analyse the effects of its measures and eco-driving,'' \emph{Sensors (Basel, Switzerland)}, vol.~14, pp. 21\,358--74, 11 2014.

\bibitem{su15010711}
F.~Karagulian, G.~Valenti, C.~Liberto, and M.~Corazza, ``A methodology to estimate functional vulnerability using floating car data,'' \emph{Sustainability}, vol.~15, no.~1, 2023.

\bibitem{doi:10.3141/2623-06}
F.~Li, K.~Tang, J.~Yao, and K.~Li, ``Real-time queue length estimation for signalized intersections using vehicle trajectory data,'' \emph{Transportation Research Record}, vol. 2623, no.~1, pp. 49--59, 2017.

\bibitem{itr2.12198}
Q.~Lin, J.~Chen, G.~Li, and Z.~He, ``Signal timing parameters inference method at intersections using license plate recognition data,'' \emph{IET Intelligent Transport Systems}, vol.~16, no.~8, pp. 1092--1107, 2022.

\bibitem{Yao31122023}
H.~W. Jiarong~Yao and K.~Tang, ``A bi-level programming method for spat estimation at fixed-time controlled intersections using license plate recognition data,'' \emph{Transportmetrica B: Transport Dynamics}, vol.~11, no.~1, pp. 1045--1070, 2023.

\bibitem{ZHAN2020102660}
X.~Zhan, R.~Li, and S.~V. Ukkusuri, ``Link-based traffic state estimation and prediction for arterial networks using license-plate recognition data,'' \emph{Transportation Research Part C: Emerging Technologies}, vol. 117, p. 102660, 2020.

\bibitem{zhouTrafficSignalPhase2024}
W.~Zhou, Y.~Wang, M.~Liu, T.~Liu, P.~Zhang, and Z.~Ma, ``Traffic {{Signal Phase}} and {{Timing Estimation Using Trajectory Data From Radar Vision Integrated Camera}},'' \emph{IEEE Transactions on Intelligent Transportation Systems}, vol.~25, no.~11, pp. 18\,279--18\,291, Nov. 2024.

\bibitem{Kerper2012}
M.~Kerper, C.~Wewetzer, A.~Sasse, and M.~Mauve, ``Learning traffic light phase schedules from velocity profiles in the cloud,'' 05 2012, pp. 1--5.

\bibitem{Axer2015}
S.~Axer and F.~Pascucci, ``Estimation of traffic signal timing data and total delay for urban intersections based on low frequency floating car data,'' 07 2015.

\bibitem{Chuang2015}
Y.-T. Chuang, C.-W. Yi, Y.-C. Tseng, C.-S. Nian, and C.-H. Ching, ``Discovering phase timing information of traffic light systems by stop-go shockwaves,'' \emph{IEEE Transactions on Mobile Computing}, vol.~14, no.~1, pp. 58--71, 2015.

\bibitem{duSignalTimingParameters2019a}
Z.~Du, X.~Yan, J.~Zhu, and W.~Sun, ``Signal {{Timing Parameters Estimation}} for {{Intersections}} using {{Floating Car Data}},'' vol. 2673, no.~6, pp. 189--201.

\bibitem{yuLearningTrafficSignal2016}
J.~Yu and P.~Lu, ``Learning traffic signal phase and timing information from low-sampling rate taxi {{GPS}} trajectories,'' vol. 110, pp. 275--292.

\bibitem{fayaziTrafficSignalPhase2015}
S.~A. Fayazi, A.~Vahidi, G.~Mahler, and A.~Winckler, ``Traffic {{Signal Phase}} and {{Timing Estimation From Low-Frequency Transit Bus Data}},'' vol.~16, no.~1, pp. 19--28.

\bibitem{7323843}
S.~A. Fayazi and A.~Vahidi, ``Crowdsourcing phase and timing of pre-timed traffic signals in the presence of queues: Algorithms and back-end system architecture,'' \emph{IEEE Transactions on Intelligent Transportation Systems}, vol.~17, no.~3, pp. 870--881, 2016.

\end{thebibliography}
\end{document}